\documentclass[a4paper,12pt]{article}

\usepackage{amsfonts}
\usepackage{graphicx}
\usepackage{amsmath}
\usepackage{color}

\newcommand{\be}{\begin{equation}}
\newcommand{\ee}{\end{equation}}

\newcommand{\f}[2]{\frac{#1}{#2}}

\def\bea{\begin{eqnarray}}
\def\eea{\end{eqnarray}}
\def\bwt{\begin{widetext}}
\def\ewt{\end{widetext}}

\begin{document}


\begin{titlepage}
\begin{center}

\noindent{{\LARGE{Conformal field theories from deformations of theories with $W_n$ symmetry}}}
\smallskip
\smallskip
\smallskip

\smallskip
\smallskip
\smallskip
\smallskip
\smallskip
\smallskip
\smallskip
\smallskip
\smallskip
\smallskip
\smallskip
\noindent{\large{Juan Pablo Babaro $^a$, Gaston Giribet $^{a,b,c}$} and Arash Ranjbar$^{b,d}$}

\smallskip
\smallskip

\end{center}
\smallskip
\smallskip

\centerline{$^a$ Departamento de F\'{\i}sica, Universidad de Buenos Aires and IFIBA-CONICET}
\centerline{{\it Ciudad Universitaria, Pabell\'on 1, 1428, Buenos Aires, Argentina.}}
\smallskip
\smallskip
\smallskip
\smallskip
\centerline{$^b$ Universit\'{e} Libre de Bruxelles and International Solvay Institutes}
\centerline{{\it ULB-Campus Plaine CPO231, B-1050 Brussels, Belgium.}}
\smallskip
\smallskip
\smallskip
\smallskip
\centerline{$^c$ Instituto de F\'{\i}sica, Pontificia Universidad Cat\'olica de Valpara\'{\i}so}
\centerline{{\it Casilla 4950, 2374631, Valpara\'{\i}so, Chile.}}
\smallskip
\smallskip
\smallskip
\smallskip
\centerline{$^d$ Centro de Estudios Cient\'{\i}ficos (CECs)}
\centerline{{\it Av. Arturo Prat 514, Valdivia, Chile.}}

\bigskip

\bigskip

\begin{abstract}

We construct a set of non-rational conformal field theories that consist of deformations of Toda field theory for ${sl}(n)$. In addition to preserving conformal invariance, the theories may still exhibit a remnant infinite-dimensional affine symmetry. The case $n=3$ is used to illustrate this phenomenon, together with further deformations that yield enhanced Kac-Moody symmetry algebras. For generic $n$ we compute $N$-point correlation functions on the Riemann sphere and show that these can be expressed in terms of $sl(n)$ Toda field theory $((N-2)n+2)$-point correlation functions. 

\end{abstract}

\end{titlepage}

\newpage



\section{Introduction}

In the last years, the interest on two-dimensional conformal field theories (CFT) with $W_n$ symmetry has been renewed, mainly because of two reasons: On the one hand, $W_n$ algebra has been identified as the charge algebra associated to the asymptotic symmetries of spin-$n$ theories in three-dimensional Anti-de Sitter spacetime \cite{Campoleoni, Henneaux}, generalizing the spin-$2$ result of Brown and Henneaux \cite{Brown-Henneaux}. On the other hand, it has been observed in \cite{AGTW} that conformal blocks of ${sl}(n)$ Toda conformal field theory, the archetypical example of a theory with $W_n$ symmetry, codify the information of the partition function of $\mathcal{N}=2$ four-dimensional superconformal quiver theories for the gauge group $SU(n)$, generalizing the Alday-Gaiotto-Tachikawa (AGT) correspondence \cite{AGT} to the case $n\geq 2$. Here, we will study the $sl(n)$ Toda field theory (TFT) from a different perspective: We will consider TFT as a generating functional of $N$-point correlation functions of a plethora of new non-rational conformal field theories that are defined ipso facto. This is in the line of the so-called $H^+_3$ Wess-Zumino-Witten -- Liouville correspondence \cite{Ribault-Teschner} or, more precisely, of the extension of it proposed in Ref. \cite{Ribault-family}, where correlation functions of a new family of non-rational CFTs were constructed in terms of Liouville field theory (namely, in terms of the $sl(2)$ Toda field theory). Here, we provide a generalization of these results to the case $sl(n)$.

The $H^+_3$ Wess-Zumino-Witten -- Liouville correspondence has been proposed in Ref. \cite{Ribault-Teschner}. It is based on a remarkable relation existing between the $sl(2)$ Knizhnik-Zamolodchikov (KZ) equation and the Belavin-Polyakov-Zamolodchikov (BPZ) equation \cite{Stoyanovsky}. The latter is the differential equation obeyed by Liouville correlation functions that involve degenerate fields, usually denoted by $V_{-1/(2b)}$. Using that solutions of the BPZ equation can be related to solutions of the KZ equation for $sl(2)$, it was shown in \cite{Ribault-Teschner} that $N$-point correlators in $H_3^+= SL(2,\mathbb{C})/SU(2)$ Wess-Zumino-Witten (WZW) on the sphere can be written in terms of $(2N-2)$-point correlators in Liouville field theory formulated also on the sphere, where the latter correlators include $N-2$ degenerate fields $V_{-1/(2b)}$. This correspondence has been later generalized to genus-$g$ \cite{Schomerus}, to surfaces with boundaries \cite{Ribault-boundary, Ribault2, Ribault3, Ribault-Hosomichi}, to spectrally flowed representations of $\hat{sl}(2,\mathbb{R})$ \cite{Ribault-winding}, to the irregular vertex representations of the Virasoro algebra \cite{Gaiotto}, and to supersymmetric theories \cite{Schomerus2}. In Ref. \cite{Ribault-family}, a generalization of the $H^+_3$ WZW--Liouville correspondence was proposed for the case in which the Liouville correlators include $N-2$ degenerate fields of higher level, $V_{-m/(2b)}$ with $m\in \mathbb{Z}_{>0}$. In the case $m=1$, these correlators yield correlators of $H^+_3$ WZW \cite{Ribault-Teschner, Stoyanovsky}; however, in the case $m>1$ the $(2N-2)$-point Liouville correlators including $N-2$ fields $V_{-m/(2b)}$ were argued to generate the $n$-point correlation functions of a new family of non-rational conformal field theories \cite{Ribault-family}. This family of CFTs is parameterized by $m$, which enters in the central charge as $c=3+6(b+(1-m)/b)^2$, with the Liouville central charge being $c_L=1+6(b+1/b)^2$. The consistency of these theories was later studied in Refs. \cite{Giribet-Babaro1, Giribet-Babaro2, Nicolas}, where correlation functions on different surfaces were computed. It is possible to speculate that such CFTs actually exist even for other real non-integer values $m$, where the Liouville correlators do not necessarily involve degenerate representations.

Another natural question that has been addressed in the literature is whether an extension of the $H^+_3$ WZW--Liouville correspondence is possible for higher rank algebras. That is, whether a relation of this sort also exists between, say, the $SL(n,\mathbb{R})$ WZW theory and the $sl(n)$ Toda theory, which would extend the results of \cite{Ribault-Teschner, Ribault-winding} to the case $n\geq 2$. This problem has been explored in Ref. \cite{Ribault-sl3} for the case $n=3$, where such a correspondence has been observed to emerge in the highly quantum limit $k\to 3$, where the WZW level equals the Coxeter number of the $sl(3)$ algebra. In the recent paper \cite{Hikida-sl3}, a more general correspondence for cases $n\geq 3$ was achieved by considering particular $sl(2)$ embeddings in (super)algebras. 

Here, aimed at investigating whether a generalization of the construction of \cite{Ribault-family} to higher rank is possible, we will propose a larger family of non-rational CFTs whose $N$-point correlation functions are given by a subset of $(N+(N-2)r)$-point correlation functions of $sl(n)$ Toda field theories, with $r=n-1$ being the rank of $sl(n)$. These new CFTs are parameterized by a collection of $n-1$ real numbers $\{ m_1,m_2,...\ m_{n-1}\} $. They correspond to deformations of Toda field theories that, while preserve confomal invariance, seem to break the $W_n$ symmetry. We will see, however, that the CFTs constructed through this procedure may also preserve a larger symmetry, generated by an affine extension of a proper subalgebra of the Borel subalgebra of the affine Kac-Moody $\hat{sl}(n)_k$. The case $n=3$ is analyzed explicitly to illustrate such remnant infinite-dimensional symmetry in relation to the symmetries of the $sl(3)$ WZW model. We will present a Lagrangian representation of these new non-rational CFTs, which is the natural generalization of the Lagrangian proposed in \cite{Ribault-family} for $n=2$, here involving interaction operators between the $n-1$ scalar fields associated to the simple roots of $sl(n)$ and $n-1$ copies of the $\beta $-$\gamma $ ghost system. In the case $n=2$ the theories coincide with those defined in Ref. \cite{Ribault-family}; that is, for $n=2$ and $m_1=1$ it corresponds to the $H_3^+$ WZW model\footnote{One may wonder whether the non-rational CFTs defined in this way are unitary. Being $c>1$ theories defined as deformation of Toda field theories, it is probable that, at least in some particular cases, the theory can be render unitary provided one considers some restriction of the Hilbert space. This is the case, for instance, of the $SL(2,\mathbb{R})$ WZW model, which represents a unitary theory when one restricts the isospin $j$ of the discrete representations.}, while for $n=2$ and $m_1=0$ it reduces to Liouville theory. In general, the {\it undeformed} case $m_i=0$ with $i=1, 2, ...\ n-1$ for arbitrary $n$ it corresponds to $sl(n)$ TFT.

The paper is organized as follows: In section 2, we briefly review the theory of Toda for the Lie algebra $sl(n)$. In section 3, we propose the Lagrangian representation of a family of CFTs which consist of deformations of TFT Lagrangian. This is a generalization of the theories proposed in \cite{Ribault-family} to the case $n>2$. We compute the correlation functions for these theories in the path integral approach, following the techniques developed in \cite{Schomerus} adapted to this case. We show that these correlation functions are determined in terms of those of TFT. We also derive an integral representation for the $N$-point functions in the Coulomb gas formalism. In section 4, we study the remnant affine symmetry that the deformed TFTs exhibit. We consider the case of $sl(3)$ to illustrate the details of such symmetry. We discuss the relation between the deformation of the $sl(3)$ TFT and the $sl(3)$ WZW model, which leads to suggest other deformations of the former. Section 5 contains our conclusions.

\section{Toda conformal field theory}

The $sl(n)$ TFT is a conformal field theory whose degrees of freedom are represented by $n-1$ bosons living in the $(n-1)$-dimensional root space of $sl(n)$ Lie algebra; see for instance Refs. \cite{Fateev-Litvinov, Fateev-Litvinov2, Miura, Fateev-Litvinov3, Fateev-Ribault}. There are $n-1$ simple roots $e_1$, $e_2$, ..., $e_{n-1}$ in this Lie algebra and its Cartan matrix is given by
\begin{equation}
K_{ij}\  =\ \left(
  \begin{array}{cccccc}
    2 & -1 & 0 & \cdots & \cdots & 0 \\
    -1 & 2 & -1 & \cdots & \cdots & \cdots \\
    0 & -1 & 2 & \cdots & \cdots & \cdots \\
    \cdots & \cdots & \cdots & \cdots & \cdots & \cdots \\
    \cdots & \cdots & \cdots & \cdots & 2 & -1 \\
    0 & \cdots & \cdots & \cdots & -1 & 2 \\
  \end{array}
\right) .
\end{equation}
This defines the inner product $(. , . )$. That is to say, this defines the bilinear form $K_{i\ j}=(e_i,e_j)$ with $K_{i\ i+1}=K_{i+1\ i}=-1$, $K_{i\ i}=2$, and $0$ otherwise in the basis above. The $n-1$ fundamental weights $\omega _i$ are defined in such a way they span dual roots, i.e. $(\omega_i,e_j)=\delta_{ij}$.

The Lagrangian of the $sl(n)$ Toda CFT is
\begin{equation}
S_T\ =\ \frac{1}{2\pi}\int d^2z\bigg((\partial\varphi,\bar{\partial}\varphi)+\frac{(Q_T,\varphi)}{4}\mathcal{R}+
\sum_{k=1}^{n-1}e^{\sqrt{2}b(e_k,\varphi)}\bigg)\label{toda}
\end{equation}
where $b$ is an arbitrary (real) parameter, and the background charge is
\begin{equation}
Q_T\ =\ \sqrt{2}\bigg( b+\frac{1}{b}\bigg)\rho \ , \label{telamorfaste}
\end{equation}
where $\rho$ is the Weyl vector; that is, $\rho $ is the half-sum of all positive roots. Normalization of the action is such that the free field propagator of the fields are 
\begin{equation}
\langle \varphi_{k }(z)\varphi_{\ell }(w)\rangle \ =\ -\frac{1}{2}\delta_{k\ell }\log (z-w) \ ,
\end{equation}
with $k, \ell = 1, 2, ...\ n-1$.

For every integer value $n$, action (\ref{toda}) defines a different theory. The case $n=2$ corresponds to Liouville field theory; in such case there is just one simple root $e_1=\sqrt{2}$ and just one highest weight $\omega_1=1/\sqrt{2}$. Liouville field theory is a conformal field theory with central charge $c=1+6Q_L^2$ and background charge $Q_L=(b+b^{-1})$. For $n>2$ the theory contains $n-1$ fields of the Liouville type interacting each other by specific couplings that are given by the $sl(n)$ structure. The resulting theory, the $sl(n)$ TFT, is also conformal invariant and its central charge is given by
\begin{equation}
c_T\ =\ n-1+6Q_T^2\ =\ (n-1)\left(1+n(n+1)\left(b+{b}^{-1}\right)^2\right),
\end{equation}
where $Q_T^2= (Q_T,Q_T)$.

These theories present spin-$n$ symmetry, represented by $n-1$ holomorphic and $n-1$ antiholomorphic conserved currents. These generate the ${ W_n}\oplus { \bar{W}_n}$ algebra, which contains Virasoro as a subalgebra (see (\ref{TToda})-(\ref{Virato}) below). Let us denote such currents by ${W}^s(z)$ and ${\bar{W}^s (\bar{z})}$ with ${s}=1,2,3,...\ n$. The index ${s}$ labels the spin of the current. These admit a representation in terms of Miura transformation \cite{Miura}
\begin{equation}
\prod_{r=0}^{n-1}\left((q+q^{-1})\partial+\sqrt{2}(\varrho_{n-r},\partial\varphi)\right)\ =\ \sum_{r=0}^n{W^{n-r}}(z)((q+q^{-1})\partial)^r
\end{equation}
where the vectors $\varrho_k$ are the weights of the first fundamental representation of the Lie algebra
$sl(n)$ with the highest weight $\omega_1$, namely $\varrho_k\ =\ \omega_1 - e_1-e_2 -...\ e_{k-1}.$

The first and second currents are taken to be ${ W^0}=1$ and ${ W^1}=0$; the third one is the first nontrivial one and corresponds to the energy-momentum tensor
\begin{equation}
{W^2}(z) =\ -\left(\partial\varphi,\partial\varphi\right)+\left(Q_T,\partial^2\varphi\right), \label{TToda}
\end{equation}
while $W^3(z)$ is the generator of the Zamolodchikov algebra \cite{Zamolodchikov}. It is customary to write these algebras (and their $n>3$ generalization) in terms of the modes of the currents, which are defined by the expansion
\begin{equation}
{W^s}(z) =  \sum_{r\in \mathbb{Z}} {W^s}_r \ z^{-r-s} \ .
\end{equation}

In terms of the modes, and in the case $n=3$, in addition to the Virasoro algebra
\begin{eqnarray}
[W_n^2 , W_m^2 ] = (n-m) W_{n+m}^2 + \frac{c}{12} n(n^2-1)\delta_{n+m , 0} \label{Virato}
\end{eqnarray}
the algebra includes the Lie products
\begin{eqnarray}
[W_n^2 , W_m^3 ] = (2n-m) W_{n+m}^3  
\end{eqnarray}
and
\begin{eqnarray}
[W_n^3 , W_m^3 ] &=& \frac{1}{15} (n-m) \left( n^2+ m^2 -\frac{1}{2} nm -4\right) W_{n+m}^2 + \nonumber \\
&&\frac{16}{22+5c}(n-m)\Lambda_{n+m}+
\frac{c}{360} n(n^2-1)(n^2-4) \delta_{n+m,0} \ , \label{coso}
\end{eqnarray}
where the quadratic piece in (\ref{coso}) is given by the normal ordered product 
\begin{equation}
\Lambda_{n} = \sum_{m\in \mathbb{Z}} :W^2_{m}W^2_{n-m}: - \frac{1}{4} t_{n}W^2_n \ , 
\end{equation}
with $t_n = n^2 -4$ for $n$ even, and $t_n = n^2+2n-15$ for $n$ odd. As usual, the normal order is defined as $:W^2_nW^2_m:\ \equiv \ W^2_nW^2_m \theta (n-m)+W^2_mW^2_n \theta (m-n)$, with $\theta (x)=\int_{\mathbb{R}_{<x}} dy\delta (y)$ being the Heaviside step function with $\theta(0)=1/2$. The OPE that realizes the $W_3$ algebra (\ref{Virato})-(\ref{coso}) takes the form
\begin{eqnarray}
W^2(z)W^2(w) &=& \frac{c/2}{(z-w)^4} + \frac{2W^2(w)}{(z-w)^2} + \frac{\partial W^2(w)}{(z-w)} + ... \nonumber \\
W^2(z)W^3(w) &=& \frac{3W^3(w)}{(z-w)^2} + \frac{\partial W^3(w)}{(z-w)} + ... \nonumber \\
W^3(z)W^3(w) &=& \frac{c/3}{(z-w)^6} + \frac{2W^2(w)}{(z-w)^4} + \frac{\partial W^2(w)}{(z-w)^3} + \frac{1}{(z-w)^2}\bigg( \frac{32}{22+5c} \Lambda (w) + \frac{3}{10} \partial^2W^2(w) \bigg) + \nonumber \\
&+&  
\frac{1}{(z-w)}\bigg( \frac{16}{22+5c} \partial \Lambda (w) + \frac{1}{15} \partial^3W^2(w) \bigg)+ ... \nonumber \\
\end{eqnarray}
with $\Lambda(z) = :(W^2(z))^2:-(3/10)\partial^2W^2(z)$. 

The basic objects in the construction of conformal TFT are exponential fields parameterized by an $(n-1)$-component vector parameter $\alpha = (\alpha_1 , \alpha_2, ...\ \alpha_{n-1})$; namely
\begin{equation}
V_{\alpha}^T(z)\ =\ e^{2(\alpha,\varphi (z))} \ ,
\end{equation}
which are in correspondence with primary states of the theory $|\alpha \rangle = \lim_{z\to 0 } V^T_{\alpha }(z)| 0 \rangle $. The operator product expansion (OPE) between the currents and these primary fields read
\begin{equation}
{W^k}(z)V_{\alpha}^T(w)\ =\ \frac{h_{\alpha}^{k}V_{\alpha}^T(w)}{(z-w)^k}\, \left( 1+ \ldots \right)
\end{equation}
where the ellipses stand for terms that vanish when $w\to z$. In particular, it yields
\begin{equation}
{W^2}(z)V_{\alpha}^T(w)\ =\ \frac{h_{\alpha}^{2}V_{\alpha}^T(w)}{(z-w)^2}+ \frac{W^{2}_{-1} V_{\alpha}^T(w)}{(z-w)}+ ... 
\end{equation}
with $W^{2}_{-1}V_{\alpha}^T(w) = \partial V_{\alpha}^T(w)$. From this, we obtain the conformal dimension
\begin{equation}
h_{\alpha}^{2} =\ \left(\alpha,Q_T-\alpha\right).
\end{equation}
The OPE with the $W^3(z)$ current reads
\begin{equation}
{W^3}(z)V_{\alpha}^T(w)\ =\ \frac{h_{\alpha}^{3}V_{\alpha}^T(w)}{(z-w)^3}+
\frac{W^{3}_{-1} V_{\alpha}^T(w)}{(z-w)^2}+ \frac{W^{3}_{-2} V_{\alpha}^T(w)}{(z-w)}+ ...  
\end{equation}
where $h_{\alpha }^3$ is an expression cubic in $\alpha $ and quadratic $b$, and symmetric under $b\leftrightarrow b^{-1}$.

The $N$-point correlation functions in TFT on the Riemann sphere are defined by
\begin{equation}
\left\langle \prod_{i=1}^N V^T_{\alpha_i}(z_i)  \right\rangle_{\text{TFT}} \ \equiv \ \int \prod_{a=1}^{n-1} {\mathcal D}\varphi _a \ e^{-S_T} \ \prod_{i=1}^{N} e^{2(\alpha_i , \varphi (z_i))}  \label{BBBB}
\end{equation}
where the fields in the functional integral are defined on $\mathbb{CP}^1 \setminus \{z_1, z_2, ... \ z_N \}$. In the next section, we will propose a definition of a set of two-dimensional CFTs whose correlation functions can be expressed in terms of TFT correlation functions (\ref{BBBB}). Proving such correspondence between observables in the case of $sl(2)$ can be assisted by the modular differential equations that correlators that involve degenerate representations obey. In the case of TFT for $sl(n>2)$, the structure of degenerate and semi-degenerate is less restrictive \cite{Ribault-sl3, Fateev-Litvinov, Fateev-Litvinov2, Miura, Fateev-Litvinov3}. Because of that, and because we are interested in defining the theories for generic values of the parameters $m_i$, in this paper we will resort to the path integral approach, following the techniques of reference \cite{Schomerus}. We will also derive an integral representation for the $N$-point functions resorting to the free field approach.

\section{Deformations of Toda field theory}

\subsection{Lagrangian representation}

Generalizing $sl(n)$ conformal Toda theories, and generalizing at the same time the family of theories presented in \cite{Ribault-family}, here we will consider the theories defined by the following actions
\begin{equation}\begin{split}
S_{\{ m_k \}}\ =\ \frac{1}{2\pi }\int  d^{2}z
\bigg( & (\partial \phi, \bar{\partial}\phi) \,+\,\sum_{k=1}^{n-1}\left(\beta_k\bar{\partial}\gamma_k\,+\,%
\bar{\beta}_k\partial \bar{\gamma}_k\right)\,+\,\frac{(Q_{\{m_k\}},\phi)}{4}\mathcal{R}\ +
\\& \,\sum_{k=1}^{n-1}(-\beta_k\bar{\beta}_k)^{m_k}\,e^{\sqrt{2}b(e_k,\phi)}\bigg)\label{nuevaaccion}
\end{split}\end{equation}
which, apart from the $(n-1)$-component boson field $\phi =(\phi_1, \phi_2, ...\ \phi_{n-1})$ multiplied with the $sl(n)$ based inner product, includes $n-1$ copies of the commutative $\beta $-$\gamma $ ghost system of conformal weight $(1,0)$. The definition of the theory also requires the specification of $n-1$ (real) parameters $m_k$, $k=1, 2, ...\ n-1$, and the parameter $b$. For each collection $\{ m_k\} =\{ m_1, m_2, ...\ m_{n-1}\} $ one obtains a different\footnote{It is, however, possible that a given CFT is represented by more than one set $\{ m_1, m_2, ...\ m_{n-1}\}$; see \cite{Nicolas}.} CFT, which is a deformation of TFT. The background charge depends on these parameters $m_k$; it is given by
\begin{equation}
Q_{\{m_k\}}=Q_T\,+\delta Q \ , \ \ \ \ \text{with}\ \ \ \delta Q =- \sum_{k=1}^{n-1}\frac{\sqrt{2}m_k}{b}\omega_k \ , \label{deformaste}
\end{equation}
where $Q_T$ is the Toda background charge (\ref{telamorfaste}). The specific value (\ref{deformaste}) is such that the interaction terms in (\ref{nuevaaccion}) are marginal. This value for the background charge yields the central charge $c_{\{m_k\}} = 3(n-1)+6Q_{\{m_k\}}^2$. This can be obtained from the OPE
\begin{equation}
W^2(z)W^2(w) = \frac{c_{\{m_k\}}/2}{(z-w)^4} + \frac{2W^2(w)}{(z-w)^2} + \frac{\partial W^2(w)}{(z-w)} + ... 
\end{equation}
with the stress tensor being
\begin{equation}
W^2(z) = \sum_{\ell = 1}^{n-1} \beta_{\ell }\partial \gamma_{\ell} - (\partial \phi , \partial \phi) + ( Q_{\{m_k\}},  \partial^2 \phi) \ .
\end{equation}
From this one can also verify that the operators $\beta_{\ell}^{m_{\ell}}\bar{\beta}_{\ell}^{m_{\ell}}e^{\sqrt{2}b(e_{\ell},\phi)}$ in (\ref{nuevaaccion}) have actually conformal weight 1. This follows from $(e_k ,\delta Q)=-\sqrt{2}m_k /b$.

We will refer to the CFTs defined by action (\ref{nuevaaccion}) as $m$-deformed CFTs. As said in the introduction, these theories are the natural generalization of the Lagrangian proposed by Ribault in Ref. \cite{Ribault-family}. In fact, in the case $n=2$ with $m_1=1$ the theory corresponds to the level $k=b^{-2}+2$ $SL(2,\mathbb{R})$ WZW model, while for $n=2$ and $m_1=0$ it reduces to Liouville theory. In general, the {\it undeformed} case $m_k=0$ for all $k=1, 2, ...\ n-1$ for arbitrary $n$ corresponds to $sl(n)$ TFT.  


We consider primary states defined by vertex operators that are exponential functions of the fields; namely
\begin{equation}
\Phi_{\alpha }(\{p_k\}|z)\ =\ {\mathcal N} \,
e^{[p,\gamma (z)]}\,e^{{2}(\alpha,\phi (z))} \label{Gmoco}
\end{equation}
where we define $[ p,\gamma ] \equiv \sum_{k=1}^{n-1} (p_k\gamma_k -\bar{p}_k \bar{\gamma }_k)$ with $\{ p_1, \ p_2, \ ... \ p_{n-1} \} $ being complex variables, and where ${\mathcal N}$ is a normalization factor, that, in principle, can depend on the momenta $\{ p_1, \ p_2, \ ... \ p_{n-1} \} $ and $\alpha $, and on the parameters of the theory $\{ m_1, \ m_2 , \ ... \ m_{n-1} \} $ and $b$.

\subsection{Correlation functions: Path integral}

The quantities we are interested to compute are the $N$-point correlation functions
\begin{equation}
\Omega\left(\{p_k^{\nu}\}, \{\alpha^{\nu}\} |\{ z_{\nu}\} \right) = 
\left\langle \prod_{\nu =1}^{N}\Phi_{\alpha_{\nu}}(\{p_k\}^{\nu} |z_{\nu
})\right\rangle_{\{ m_k \} } \equiv \ \int \prod_{k=1}^{n-1} \mathcal{D}\phi_k \mathcal{D}%
^{2}\beta_k \mathcal{D}^{2}\gamma_k\,e^{-S_{\{m_k\}}}\prod_{\nu
=1}^{N}\Phi_{\alpha_{\nu }}(\{p_k\}^{\nu }|z_{\nu }), \label{Cosillas}
\end{equation}
where our notation is such that the symbol $\{ p_k^{\nu}\}$ in the argument of a function represents the collection of $N(n-1)$ elements $\{p_1^1, p_2^1, ... \ p_{n-1}^1 ; p_1^2, p_2^2, ... \ p_{n-1}^2 ; ...\ p_1^N, p_2^N, ... \ p_{n-1}^N \}$ on which the function depends, while the symbol $\{ p_k\}^{\nu}$ represents the collection of $(n-1)$ variables $\{p_1^{\nu }, p_2^{\nu }, ... \ p_{n-1}^{\nu }  \}$ with $\nu $ fixed. Indices $\mu , \nu$ will be used to run over $\{ 1,2, ...\ N \}$, while indices $k,\ell $ run over $\{ 1,2, ...\ n-1 \}$, and indices $a,b$ run over $\{ 1,2, ...\ N-2 \}$.

The first step in computing (\ref{Cosillas}) is dealing with the ghost system: The integration over the fields $\gamma_k $ and $\bar{\gamma}_k$ produces a product of $n-1$ double $\delta$-functions
\begin{equation}
\delta^2\left(\bar{\partial}\beta_k(w)\,-\,2\pi\sum_{\nu=1}^Np_k^{\nu}\delta^2(w-z_{\nu})\right), \label{Gdelta}
\end{equation}
which set the conditions
\begin{eqnarray}
\bar{\partial}\beta_k(z) -2\pi \sum_{\nu =1}^{N}p_k^{\nu }\,\delta
^{2}(z-z_{\nu }) =0, \ \ \ \ 
\partial \bar{\beta}_k(\bar{z})+2\pi \sum_{\nu =1}^{N}\bar{p}_k^{\nu
}\,\delta ^{2}({z}-{z}_{\nu }) =0 ,
\end{eqnarray}
for each $k$.

These $2(n-1)$ equations have solution only if the sum of the momenta $\{ p_k\}^{\nu}$ vanishes. More precisely, being meromorphic on the Riemann sphere, the sum of the residues of $\beta_k(z)$ vanishes, and therefore
\begin{equation}
\sum_{\nu=1}^{N}p_k^{\nu}=0. \label{U37}
\end{equation}

In order to write the $\beta_k(z)$ in terms of its residues $\{ p_k^1$, $p_k^2$, ... $p_k^N \}$, one may resort to the representation $\bar{\partial}({1}/{z}) =\partial({1}/{\bar{z}})= 2\pi\delta^2(z)$ and integrate. Since $\beta_k(z)$ are 1-differentials, the general solution can be written as a rational function; namely
\begin{equation}\label{U37b}
\beta_k(z)\ =\ \sum_{\nu=1}^N\frac{p_k^{\nu}}{z-z_{\nu}}\ \equiv\  \frac{{\mathcal P}_k(z,\{ y_a\}^{k} )}{{\mathcal Q}(z,\{ z_{\nu } \} )},
\end{equation}
with ${\mathcal P}_k(z,\{ y_a\}^{k} )$ being $n-1$ polynomials of degree $N-2$ in $z$, and ${\mathcal Q}(z,\{ z\}_{\mu } )$ being a polynomial of degree $N$ in $z$. In fact, for a meromorphic 1-differential $\beta_k(z)$ on the Riemann sphere, the difference between the amount of its poles $\{y_a \}^k$ ($a=1, 2, ... \ $) and the amount of its zeroes $\{z_{\mu} \}$ ($\mu=1, 2, ... \ $) is $2$. Therefore, these polynomials have the form
\begin{eqnarray}
{\mathcal P}_k(z,\{ y_a\}^{k} ) =  \kappa_k \prod_{a=1}^{N-2}(z-y^k_a) \ , \ \ \ 
{\mathcal Q}(z,\{ z_{\nu } \} )  = \prod_{\nu =1}^N (z-z_{\nu}).
\end{eqnarray}

From this, using (\ref{U37}) it is easy to show that $\kappa_k = \sum_{\nu = 1}^{N}p_{\nu }^{k}z_{\nu}$. This follows from multiplying (\ref{U37}) by ${\mathcal Q}(z,\{ z_{\nu } \} ) $ to obtain an identity between polynomials and then matching the coefficients of powers of $z$. More precisely, one finds 
\begin{eqnarray}
{\mathcal P}_k(z,\{ y_a\}^{k} )  = \sum_{\nu =1}^{N} \prod_{\mu \neq \nu }^{N} p_{k}^{\nu } (z-z_{\mu }) \equiv \sum_{n =1}^{N-1} c_{n}\ z^{N-n} \ ,
\end{eqnarray}
which yields
\begin{eqnarray}
0=c_{1} = \sum_{\mu=1}^{N} p_{k}^{\mu }  \ ,\ \ \ \ \ \ \   \kappa_k = c_{2} = -\sum_{\mu \neq \nu }^{N} p_{k}^{\nu }z_{\mu } = \sum_{\mu=1}^{N} z_{\mu } p_{k}^{\mu } .
\end{eqnarray}

It is easy to keep track of the dependence on $\kappa_k$. This is gathered by an overall factor $\prod_{k=1}^{n-1}|\kappa_{k} |^{2\lambda_k}$ in the final result. This can be seen by shifting the fields as $\phi \to \phi -\sum_{k=1}^{n-1}\sqrt{2}b^{-1}m_{k} \omega_{k} \log \kappa_{k}$ to absorb other dependence of $\kappa_k$. Then, we can omit the explicit dependence on $\kappa_k$ and restore it in the final result (\ref{relacion}).

By evaluating the residues, one finds
\begin{equation}
p_k^{\nu}\ =\ \kappa_{k} \frac{\prod_{a=1}^{N-2}(z_{\nu}-y^k_a)}{\prod_{\mu\neq\nu}^N
(z_{\nu }-z_{\mu })}. \label{Gusala}
\end{equation}

Taking all this into account, the $\delta $-function (\ref{Gdelta}) can be replaced by
\begin{equation}
\delta^2\left(\sum_{\nu=1}^Np_k^{\nu}\right)\,\prod_{k=1}^{n-1}\delta^2\left(\beta_k -
\frac{{\mathcal P}_k(z,\{ y_a\}^{k} )}{{\mathcal Q}(z,\{ z_{\mu } \} )}
\right).
\end{equation}
provided the determinant factor ${\det ^{1-n} \square }$ is trivial on the sphere (with $\square = \partial\bar{\partial}$). Then, one can integrate over $\beta_k $ and $\bar{\beta}_k$ to obtain
\begin{equation}\begin{split}
\Omega\left(\{p_k^{\nu}\}, \{\alpha_{\nu }\}|\{ z_{\nu}\} \right)\ =\ \int\prod_{k=1}^{n-1}\mathcal{D}\phi_k \,e^{-S_{\text{eff}}} \ \prod_{\nu=1}^N{\mathcal N}_{\nu }\ e^{{2}(\alpha^{\nu},\phi (z_{\nu }))}. \label{Gfinal}
\end{split}\end{equation}%
with the effective action
\begin{equation}
S_{\text{eff}} = -\frac{1}{2\pi}\int d^2z \bigg((\partial\phi,
\bar{\partial}\phi)+\frac{(Q_{\{m_k\}},\phi)}{4}\mathcal{R}\,+\,b^2\sum_{k=1}^{n-1} { | {\mathcal P}_k(z,\{ y_a\}^{k} ) /  {\mathcal Q}(z,\{ z\}_{\mu } )} |^{2m_k}
e^{\sqrt{2}b(e_k,\phi)}\bigg) , \label{Gfinal2}
\end{equation}
which depends on the $N$ inserting points $\{ z_{\mu } \}$ and the $(n-1)(N-2)$ variables $\{ y_{a}^{k}\}$, although the latter are determined by the $(n-1)N$ algebraic equations (\ref{Gusala}) with the $n-1$ constraints (\ref{U37}).

For further purpose, in order to achieve a regularization of divergence coming from coincident points in the path integral, it is convenient to write the action in the conformal frame $ds^2\ =\ e^{2\sigma }dz\,d\bar{z}$
which yield $({1}/{4})\sqrt{g}\mathcal{R} = -2 \square \sigma $. This will enable us to regularize the propagator $\langle \phi (z)\phi(w) \rangle $ in the limit $z\to w$.

As previously said, the normalization of the vertices ${\mathcal N}$ in (\ref{Gmoco}) can in principle depend on the momenta $\{ p_1, \ p_2, \ ... \ p_{n-1} \} $ and $\alpha $. For instance, the natural generalization of the normalization considered in \cite{Ribault-family} to the $sl(n\geq 2)$ case would be
\begin{equation}
{\mathcal N_{\nu}} = \prod_{k=1}^{n-1}|p^{\nu }_k|^{\frac{2m_k}{b}(\alpha^{\nu},\omega_k)}, \label{Cucote}
\end{equation}
which, after $\delta$-function evaluation and considering (\ref{Gusala}), would produce in (\ref{Gfinal}) a factor 
\begin{equation}
\prod_{\nu =1}^{N}\prod_{k=1}^{n-1}\bigg| \frac{{\mathcal P}_k(z_{\nu },\{ y_a\}^{k} )}{{\mathcal Q}(z_{\nu } ,\{ z_{\mu } \} )}\bigg|^{\frac{2m_k}{b}(\alpha_{\nu} ,\omega_k )}\ 
\end{equation}
accompanying the $N$ vertices. On the other hand, a convenient normalization also seems to be demanding ${\mathcal N}_{\nu }$ not to depend on the momenta at all. In order to consider a more general case that, in particular, includes (\ref{Cucote}), we might consider ${\mathcal N_{\nu}} = \prod_{k=1}^{n-1}|p^{\nu }_k|^{\frac{2m_k}{b}(\alpha^{\nu},\omega_k)(1-t)}$ with $0\leq t \leq 1$. This would affect the final result only in the exponent of the factor that we call $\Theta $ in (\ref{Tuto}) below.

In order to absorb the factors $| {\mathcal P}_k(z,\{ y_a\}^{k} ) / {\mathcal Q}(z,\{ z\}_{\mu } ) |^{2m_k}$ in the terms in the effective action (\ref{Gfinal2}), it is convenient to shift the scalar field $\phi $ -- which actually is a vector in the space generated by the simple roots $\{ e_{\ell } \}$ ($\ell =1, 2, ... n-1$) -- by defining the new field $\varphi $
\begin{equation}
\varphi (z)\ \equiv \phi (z) \,+\,\sum_{k=1}^{n-1}\omega_k\,\frac{\sqrt{2}m_k}{b}\left(
\log \vert  {  {\mathcal P}_k(z,\{ y_a\}^{k} )  } /{  {\mathcal Q}(z,\{ z_{\nu }\} )}\vert -\,\sigma \right). \label{Gshifting}
\end{equation}
That is
\begin{equation}
\phi (z)\ = \varphi (z)\,-\,\sum_{k=1}^{n-1}\omega_k\,\frac{\sqrt{2}m_k}{b}\left(\sum_{a=1}^{N-2}\log|z-y^k_a|^2\,-\,
\sum_{\nu=1}^N\log|z-z_{\nu}|^2\,-\,\sigma \right). \label{definito}
\end{equation}

Indeed, this produces the rescaling in the exponential potential terms, yielding
\begin{equation}
e^{\sqrt{2}b(e_k,\phi(z))}\ = \ e^{\sqrt{2}b(e_k,\varphi(z))} 
\bigg|  \frac{  {\mathcal Q}(z,\{ z_{\nu } \} ) }{ {\mathcal P}_k(z,\{ y_a\}^{k} )  }\bigg|^{2m_k} e^{2m_k \sigma } \ ,
\end{equation}
which cancels the factors $| {\mathcal P}_k(z,\{ y_a\}^{k} ) / {\mathcal Q}(z,\{ z\}_{\mu } ) |^{2m_k}$. It also produces the rescaling
\begin{equation}
{\mathcal N_{\nu}}\ e^{\sqrt{2}(\alpha_{\nu},\phi(z_{\nu}))}\ = \ e^{\sqrt{2}(\alpha_{\nu},\varphi(z_{\nu}))}
\prod_{k=1}
^{n-1}
\bigg|  \frac{  {\mathcal Q}(z,\{ z_{\mu }\neq z_{\nu } \} ) }{ {\mathcal P}_k(z,\{ y_a\}^{k} )  }\bigg|^{\frac{2tm_k}{b} (\alpha_{\nu} , \omega_k )},\label{moCo}
\end{equation}
where the divergences coming from coincident points is regularized by defining the finite part of $\underset{z_{\mu}\rightarrow z_{\nu}}{\lim}\log|z_{\mu }-z_{\nu}|^2 $ absorbing the divergence in the exponent of the conformal factor $ -2\sigma $, \cite{Polyakovo}. Let us consider the normalization (\ref{Cucote}), corresponding to $t=0$. 

Definition (\ref{definito}) also produces a shifting in the kinetic term; namely
\begin{equation}
(\phi,\square \phi)\, = \,(\varphi,\square \varphi)\,-\sum_{k=1}^{n-1}\frac{m_k}{\sqrt{2}b}(\omega_k,a_k) + 
\sum_{j=1}^{n-1}\sum_{k=1}^{n-1} \frac{m_j m_k}{2b^2} (\omega _j , b_{jk}) 
\end{equation}
with
\begin{eqnarray}
a_k(z, \{z_{\mu}\}, \{y_{a}\}^{k}) &=&
\varphi (z)\ \left(2\pi\sum_{a=1}^{N-2} \delta^2(z-y_a^k)-2\pi \sum_{\nu=1}^N\delta^2(z-z_{\nu})- 2\square \sigma \right) + \nonumber \\
&&
 \ \partial\bar{\partial}\varphi (z) \ \left(\sum_{a=1}^{N-2}\log|z-y_a^k|^2-\sum_{\nu=1}^N
\log|z-z_{\nu}|^2-2\sigma  \right) \label{GasA}
\end{eqnarray}
and
\begin{eqnarray}
b_{jk}(z, \{z_{\mu}\}, \{y_{a}\}^{k}, \{y_{a}\}^{j}) &=& 2\pi\omega_k  \ \left(\sum_{a=1}^{N-2}\log|z-y_a^k|^2-\sum_{\mu=1}^N\log|z-z_{\mu}|^2-2\sigma  \right)\,\times \nonumber \\
&&\times \left(\sum_{b=1}^{N-2}\delta^2(z-y_b^j)-\sum_{\nu=1}^N\delta^2(z-z_{\nu})\right) - \label{GasB} \\
&& 2\omega_k \left(\sum_{a=1}^{N-2}\log|z-y_a^k|^2-
\sum_{\nu=1}^N\log|z-z_{\nu}|^2-2\sigma \right) \square \sigma . \nonumber
\end{eqnarray}

Considering again the regularization of coincident points, and replacing the last equation into the effective action (\ref{Gfinal2}), one obtains the following contributions in the correlation function: From the first two terms (\ref{GasA}), one obtains
\begin{equation}
\prod_{k=1}^{n-1}\prod_{i=1}^{N-2}e^{-\frac{\sqrt{2}m_k}{b}(\omega_k,\varphi(y_i^k))} \times \prod_{\nu=1}^Ne^{\frac{\sqrt{2}}{b}\sum_{k=1}^{n-1}m_k(\omega_k,\varphi(z_{\nu}))}.
\end{equation}
From the third term, and taking into account $({1}/{4})\sqrt{g}\mathcal{R} = -2 \square \sigma$, one observes that the background charge term in the action gets shifted as $Q_{\{m_k\}}\to Q_{\{m_k\}}-\delta Q = Q_T$. The terms (\ref{GasB}), after a proper regularization of coincident insertions, yields the factor 
\begin{eqnarray}
\vert \Theta (\{ m_k \} | \{ z_{\mu } \} , \{ y_a^k \} ) \vert &=& \prod_{a=1}^{N-2} \prod_{b=1}^{N-1} \prod_{k=1}^{n-1} \prod_{l=1}^{n-1} |y_{a}^{k}-y_{b}^{l}|^{\frac{2m_k m_l}{b^2}(\omega_k , \omega_l )} 
\prod_{\mu < \nu }^{N} 
|z_{\mu }-z_{\nu }|^{\frac{2}{b^2}\sum_{k=1}^{n-1}\sum_{l=1}^{n-1}m_k m_l(\omega_k , \omega_l )}  \times \nonumber \\
&& \times \prod_{a=1}^{N-2} \prod_{k=1}^{n-1} \prod_{\mu =1}^{N}
|y_{a}^{k}-z_{\mu }|^{-\frac{2m_k}{b^2}\sum_{l=1}^{n-1}m_l(\omega_k , \omega_l )} \ . \label{Tuto}
\end{eqnarray}

Notice that this contribution corresponds exactly to the correlator of exponential vertices of a free boson $\chi $ with a background charge \cite{Giribeto}; namely 
\begin{equation}
\Theta (\{ m_k \} | \{ z_{\mu } \} , \{ y_a^k \} ) = \left\langle \prod_{\mu =1}^{N} V_{ -\sum_{k=1}^{n-1} q_k}(z_{\mu }) \prod_{a=1}^{N-2}\prod_{l=1}^{n-1} V_{q_l}(y_{a }^l) \right\rangle_{\chi }
\end{equation}
where
\begin{equation}
V_{P }(z) \ = \ e^{2iP \chi (z)} , 
\end{equation}
and where $q_{\ell }=m_{\ell }\omega_{\ell }/(\sqrt{2}b)$, Wick contracted the free field propagator $\langle \chi (z) \chi (w) \rangle \ = \ -(1/2) \log (z-w)$. The value of the background charge $Q_{\chi }$ associated to the field $\chi (z)$ can be read from the conservation law that would follow from the integration over the zero-mode $\langle \chi \rangle $. This yields the result $Q_{\chi }=\delta Q$. 

Therefore, we finally arrive to the conclusion that the $N$-point correlation functions (\ref{Cosillas}) take the form
\begin{equation}\begin{split}
\left\langle \prod_{\nu =1}^{N}\Phi_{\alpha_{\nu}}(\mu _{\nu }|z_{\nu})\right\rangle _{\{m_k \} } = 
\ \prod_{k=1}^{n-1}|\kappa_{k} |^{2 \lambda_k} \prod_{k=1}^{n-1} & \delta\left(\sum_{\nu=1}^Np_k^{\nu}\right)\,
\left\langle \prod_{\mu =1}^{N} V_{-\sum_{k=1}^{n-1} q_k}(z_{\mu }) \prod_{a=1}^{N-2}\prod_{l=1}^{n-1} V_{q_l}(y_{a }^l) \right\rangle_{\chi } \,\times\\
&\times\, \left\langle\prod_{\nu=1}^NV_{\alpha_{\nu}+\sum_{k=1}
^{n-1}q_k}^T(z_{\nu})\prod_{l=1}^{n-1}\prod_{a=1}^{N-2}V_{-q_l}^T (y^l_a)\right
\rangle_{\text{TFT}}\label{relacion}
\end{split}\end{equation}
where $\lambda_k = m_k (1+b^{-2}-\sum_{j=1}^{n-1}m_jb^{-2}(\omega_k , \omega_j ))$; notice that here we have reintroduced the dependence on $\kappa_k$. In other words, $N$-point correlation functions $\Omega (\{p^{\mu}_k\},\{\alpha_{\mu}\}|\{z_{\mu}\})$ of the theories defined by the Lagrangian representation (\ref{nuevaaccion}) turn out to be given by $((N-2)n+2)$-point correlation functions of the CFT given by the product of $sl(n)$ TFT times a free $U(1)$ boson. Having the expression (\ref{relacion}) for the correlation functions $\Omega (\{p_k^{\nu}\}, \{\alpha ^{\nu}\} |\{z_{\nu }\})$, the $m$-deformed CFTs (\ref{nuevaaccion}) are defined {ipso facto}. In the next section, we will analyze the symmetries of the new theories.

\subsection{Correlation functions: Coulomb gas}

Lagrangian representation (\ref{nuevaaccion}) enables us to compute correlation functions in the so-called Coulomb gas approach and thus provide an integral representation of $N$-point correlation functions. This amounts to consider the expectation values (\ref{Cosillas}) and first integrate over the zero-modes of the fields $\phi_k$. This yields
\begin{equation}
\Omega\left(\{p_k^{\nu}\}, \{\alpha^{\nu}\} |\{ z_{\nu}\} \right) = \prod_{l =1}^{n-1} (-1)^{m_l s_l}\Gamma(-s_l) \int \prod_{k=1}^{n-1} \prod_{r =1}^{s_k} d^2w_{r}^{k}
\ \left\langle \prod_{\nu =1}^{N}\Phi_{\alpha_{\nu}}(\{p_k\}^{\nu} |z_{\nu
}) \prod^{n-1}_{k=1} \prod_{r = 1}^{s_k} S^{k}(w^k_{r})
\right\rangle_{\text{free} } \nonumber
\end{equation}
where $\{w_{r}^{k}\} = \{ w_{1}^{1}, w_{2}^{1}, ...\ w_{s_1}^{1};w_{1}^{2}, w_{2}^{2},...\ w_{s_2}^{2}; ... \ ;w_{1}^{{n-1}}, w_{2}^{{n-1}}, ... w_{s_{n-1}}^{{n-1}} \}$ are complex variables, where the screening operators $S^{k}(w)$ are given by
\begin{equation}
S^{k}(w)\equiv \beta^{m_{k}}{(w)}\bar{\beta}^{m_{k}}{(\bar{w})} e^{\sqrt{2}b(e_k , \phi(w))} \ ,
\end{equation}
and where the expectation value $\langle ... \rangle _{\text{free}}$ is defined in terms of the free action
\begin{equation}\begin{split}
S_{\text{free}}\ =\ \frac{1}{2\pi }\int  d^{2}z
\bigg(  (\partial \phi, \bar{\partial}\phi) \,+\,\sum_{k=1}^{n-1}\left(\beta_k\bar{\partial}\gamma_k\,+\,%
\bar{\beta}_k\partial \bar{\gamma}_k\right)\,+\,\frac{(Q_{\{m_k\}},\phi)}{4}\mathcal{R} \bigg) \ . \label{Fret}
\end{split}\end{equation}

The dependence on $\{m_k\}$ is through the number of integrals to be performed, and it is given by
\begin{equation}
s_k = {\sqrt{2}}{b}^{-1}( \Delta , \omega_k) \ , \ \ \ \ \Delta = \sqrt{2} (b+b^{-1})\rho -\sqrt{2}b^{-1}\sum_{l=1}^{n-1} m_l\omega_l  -\sum_{i=1}^{N} \alpha_i  . \label{conference52}
\end{equation}
This equation, as well as the $\Gamma (-s_l)$ factors above, comes from the $\delta$-functions arising in the integration over the zero-modes $\langle \phi_k \rangle$, what selects in the path integral the terms satisfying the vector equation
\begin{equation}
\sqrt{2}\sum_{i=1}^{N} \alpha_i + \sum_{k=1}^{n-1} (s_k e_k b + {2}b^{-1}m_k \omega _k) -{2} (b+b^{-1})\rho = 0 .
\end{equation}
The integrals are over the complex hyperplane $\mathbb{C}^{\sum_{l=1}^{n-1}s_l}$, where $w_{r}^{k}$ represent $s_1 + s_2 + ...\ s_{n-1}$ complex integration variables, with $k=1$,$2$,$...$ $n-1$ and $r=1$,$2$,$...$ $s_k$. 

The fact that the expectation value is now defined in terms of action (\ref{Fret}) enables one to use the free field propagators. In particular, one has the Coulomb propagator $\langle \phi_{\ell} (z_i)\phi_k (z_j)\rangle = -(\delta_{\ell , k}/2)\log (z_i -z_i)$. For the set of correlators whose vertices obey the kinematic condition $p_k^{\mu}=\bar{p}^{\mu}_k=0$ for all $k$ such that $m_k\neq 0$ and for all $\mu =1, 2, ... \ N$, one eventually finds
\begin{eqnarray} 
\Omega\left(\{p_k^{\nu}\}, \{\alpha^{\nu}\} |\{ z_{\nu}\} \right) = \prod_{\mu =1}^N{\mathcal N}_{\mu } \ \prod_{l=1}^{n-1} (-1)^{s_lm_l}\Gamma(-s_l) \prod_{\mu <\nu }^{N} |z_{\mu } -z_{\nu }|^{-4(\alpha _{\mu } , \alpha_{\nu } )}\ I_N(\{\alpha_{\mu}\},\{m_k\}|\{z_{\mu}\}) \nonumber
\end{eqnarray}
with
\begin{eqnarray}
I_N(\{\alpha_{\mu}\},\{m_k\}|\{z_{\mu}\})&=&\int \prod_{k=1}^{n-1}\prod_{r =1}^{s_k} d^2w_{r}^{k} \prod_{k=1}^{n-1} \prod_{t=1}^{s_k} \prod_{l=1}^{s_{k+1}}  | w^k_{t} - w^{k+1}_{l} |^{2b^2}  \prod_{k=1}^{n-1} \prod_{r < t}^{s_k}   |w^k_{r} -w^k_{t} |^{-4b^2}    \times \nonumber \\
&&
\times \prod_{\mu =1}^{N} \prod_{k=1}^{n-1} \prod_{r= 1}^{s_k}    |w^k_{r}-z_{\mu }|^{-2\sqrt{2}b(e_{k},\alpha_{\mu })}
\label{integrator}
\end{eqnarray}
where $K_{ij}=(e_i , e_j)$ has been used. In the cases in which for some value of $k$ it happens that $p_k\neq 0\neq m_k$, then a similar representation exists, although the combinatorics when performing the Wick contraction is more involved because of the $\beta $-$\gamma$ systems, which in particular yield $\beta_i (w)e^{[p,\gamma (z)]}\simeq p_i e^{[p,\gamma (w)]}/(w-z)+...$. This is similar to the contraction between string theory tachyon and graviton vertices. The Wick contraction for these operators in the case $n=2$ is discussed in \cite{Schomerus}. The same can be applied here. For instance, in the case of the 2-point function the Wick contraction of the holomorphic piece of the $\beta$-$\gamma$ system yields a factor
\begin{equation}
\prod_{k=1}^{n-1} \bigg( \prod_{i=1}^{2}\prod_{r=1}^{s_{k}} (z_i - w_r^k)^{-1} \bigg( \prod_{l=1}^{s_k} (-w_l^k) (p_1+p_2)^{s_k} + ... \ \bigg) \bigg)
\end{equation}
where the ellipses represent contributions with less $w$'s factors. Using (\ref{U37}), one observes that, eventually, only a contribution proportional to 
\begin{equation}
\propto \prod_{k=1}^{n-1}\prod_{i=1}^{2}\prod_{r=1}^{s_{k}} |z_i - w_r^k|^{-2} \label{tormentita}
\end{equation}
survives. This produces an additional shift in the exponent in the last factor of (\ref{integrator}).

Integral representation (\ref{integrator}), which is similar to the one that originally appears in the context of Minimal Models \cite{DF}, can be solved in some very special cases using the techniques of Refs. \cite{Fateev-Litvinov, Fateev-Litvinov2}. This yields closed expressions for reflection coefficients, structure constants, and spherical partition function in several CFTs, including non-rational ones \cite{GN3}. Specially in the latter case, an analytic continuation is required as the expression (\ref{integrator}) only makes sense for $s_k\in \mathbb{Z}_{\geq 0}$. Such extension for values $s_k \in \mathbb{C}$ has been successfully carried out in diverse examples, including non-compact timelike CFTs \cite{Gasto}. 

Notice that, even when the integral representation (\ref{integrator}) looks very similar to the one that would appear in the analogous computation for TFT, the integrals appearing in both cases are not exactly the same, one of the differences being the amount of integrations to be performed: While for the $m$-deformed theory one finds (\ref{conference52}), the analogous quantity in an $M$-point TFT correlation function requires not $s_k$ but
\begin{equation}
s_k^T = s_k+ 2b^{-2} \sum_{l=1}^{n-1}m_l (\omega_l , \omega_k) = -\sqrt{2}\sum_{i=1}^{M}(\alpha^T_i ,\omega_k ) + 2(b+b^{-1})(\rho ,\omega _k) \label{tormentita2}
\end{equation}
integrals. This is consistent with the shifting in the momenta $\{ \alpha _i \}$ and the presence of additional $M-N=(N-2)(n-1)$ vertices in the TFT correlator on the right hand side of equation (\ref{relacion}), as well as with the shifting $\delta Q$ in the background charge. To see this in a concise example, let us compute the 2-point functions in the theory (\ref{nuevaaccion}) using the integral representation above. That is, let us compute 
\begin{equation}
\langle \Phi_{\alpha } (\{p_k\}^1|z_1) \Phi_{\alpha } (\{p_k\}^2|z_2) \rangle _{\{m_k\}} = \frac{R(\alpha )}{|z_1-z_2|^{4\tilde{h}^2_{\alpha}}},
\end{equation}
where $\tilde{h}^2_{\alpha}= (\alpha , Q+\delta Q - \alpha )$ is the conformal dimension. The relevant information here is given by the reflection coefficient $R(\alpha)$. Conformal invariance permits to fix three vertex insertions on the Riemann sphere; as usual, let us fix the inserting points of the vertex operators $\Phi_{\alpha } (\{p_k\}^i|z_i)$ at $z_1=0$ and $z_2=1$, together with one screening operator at $\infty $. Taking into account (\ref{tormentita}), one observes that (\ref{tormentita2}) maps into (\ref{conference52}) if one identifies the TFT momenta $\alpha_i^T$ as follows
\begin{equation}
\alpha_i^T = \alpha_i + \sum_{l=1}^{n-1} q_l , \label{shiftedmomentum}
\end{equation}
which turns out to be in perfect agreement with (\ref{relacion}) for $N=2$. On the one hand, this shows that the Coulomb gas realization above is consistent with the relation between correlators given in (\ref{relacion}). On the other hand, this gives the expression for the reflection coefficients of the CFTs defined by (\ref{nuevaaccion}), which turns out to be given by the TFT analogous quantity \cite{FateevR} evaluated in the shifted momentum (\ref{shiftedmomentum}). The explicit form of TFT reflection coefficient is such that making the replacing (\ref{shiftedmomentum}) results in the shifting $Q_T\to Q$, as expected. This is because the TFT reflection coefficient depends on the momentum\footnote{See Eqs. (1.14)-(1.17) of Ref. \cite{FateevR}. Our convention relates to the one there by performing the changes $\phi \to \phi / \sqrt{2} $, $\alpha^T \to \alpha / \sqrt{2} $, and $Q_T \to \sqrt{2} Q $.} through the combination $2\alpha^T - Q_T$.

\section{Remnant affine symmetry}

\subsection{Remnant symmetry}

Let us consider the case $n=3$ with deformation parameters $m_1$ and $m_2$. It turns out that, remarkably, in that case the theory (\ref{nuevaaccion}) results to be invariant under the symmetry generated by the currents
\begin{eqnarray}
J_1^+ (z) &=& {m_1}^{-1} \beta_1   \ , \ \ \ \ \   J_1^0 (z) = {\sqrt{2}m_1}{b}^{-1}(e_1,\partial\phi)-{2\gamma_1\beta_1}+{m_1}{m_2}^{-1}\gamma_2\beta_2   \nonumber \\
J_2^+ (z) &=& {m_2}^{-1} \beta_2  \ , \ \ \ \ \   J_2^0 (z) = {\sqrt{2}m_2}{b}^{-1}(e_2,\partial\phi)+{m_2}{m_1}^{-1}\gamma_1\beta_1-{2}\gamma_2\beta_2 
\label{nuevascorrientes}
\end{eqnarray}
and their anti-holomorphic counterparts $\bar{J}^A_i$ ($A=0,\pm $, $i=1,2$), with the free field correlators
\begin{equation}
\langle \phi _i (z) \phi_{j} (w)\rangle = -\frac{1}{2}\delta_{ij } \log (z-w) \ ,  \ \ \ \ \langle \beta_i (z) \gamma_j (w) \rangle = \frac{\delta_{ij }}{(z-w)} 
\end{equation}
with $i,j\ = 1,2 $. It is possible to verify that the OPE between the interaction term of the action (\ref{nuevaaccion}) and the currents (\ref{nuevascorrientes}) has no singular term up to a total derivatives.

The symmetry algebra is encoded in the singular terms of the OPE between the currents (\ref{nuevascorrientes}). The non-regular OPEs read 
\begin{equation}
J_i^+(z)J_j^0(w)\ \sim\ \frac{(3\delta_{ij}-1)}{(z-w)}\ J_i^+(w) \,+\ ... ,\ \ \
J_i^0(z)J_j^0(w)\ \sim\ -\frac{c_{ij}}{(z-w)^2}\,+\ ... ,\nonumber
\end{equation}
and regular otherwise. The coefficients of the central terms of the algebra are given by
\begin{equation}
c_{ij}= m_i m_jb^{-2} (e_i , e_j ) + 4\delta_{ij} -2(m_i^2 + m_j^2){m^{-1}_i m^{-1}_j} |\varepsilon_{ij}| + m_i m_j \sum_{k} m_k^{-2} |\varepsilon_{ik} \varepsilon_{jk} |  
\end{equation} 
where $\varepsilon_{12}=-\varepsilon_{21}=1$, $\varepsilon_{11} = \varepsilon_{22} = 0$, and $i,j,k=1,2$. Some of these coefficients, however, can be changed by changing the normalization of the currents. In terms of the modes $J_{i,n}^a$, which are defined by $J_{i}^a(z) = \sum_{n\in \mathbb{Z}} J_{i,n}^a\ z^{-n-1}$ (with $i=1,2$ and $a=0,+$), the symmetry algebra reads $[J_{i,n}^{0},J_{j,m}^{0}]=(n/2)c_{ij} \delta_{n+m,0}$, $[J_{i,n}^{+},J_{i,m}^{0}]= 2 J_{i, n+m}^{+}$, $[J_{i,n}^{+},J_{i\neq j,m}^{0}]= - J_{i, n+m}^{+}$, $[J_{i,n}^{+},J_{j,m}^{+}]= 0$.

This can be extended to the $sl(n)$ case, which one can actually verify to be symmetric under the $2n-2$ currents  
\begin{eqnarray}
J_k^+ (z)= {m_k}^{-1} \beta_k   \ , \ \ \ \ \   J_k^0 (z)= {\sqrt{2}m_k}{b}^{-1}(e_k,\partial\phi)-{3\gamma_k\beta_k}+{m_k}\sum_{l=k-1}^{k+1}{m_l}^{-1}\gamma_l\beta_l   
\label{nuevascoientes}
\end{eqnarray}
with $k,l=1,2, ... n-1$ and where $m^{-1}_{0}=m^{-1}_{n}=0$.

Let us denote by $\hat{\mathcal A}_n \oplus \hat{\mathcal A}_n$ the algebra generated by (\ref{nuevascoientes}) and by its anti-holomorphic counterparts. Some properties of this algebra are the following: Algebra $\hat{\mathcal{A}}_n$ is the affine extension of the semi-direct sum of two Abelian Lie algebras ${\mathcal{A}}_n^{+}$ and ${\mathcal{A}}_n^{0}$; that is, $\hat{\mathcal{A}}_n = \hat{\mathcal{A}}_n^+ \oplus _s \hat{\mathcal{A}}_n^0$ with $J^+_k (z)$ and $J^0_k (z)$ generating each of the two pieces respectively. Algebras $\mathcal{A}_n^+$ and $\mathcal{A}_n^0$ are Abelian and of dimension $n-1$. $\mathcal{A}_n^0$ is the Cartan subalgebra of $sl(n)$. While $\hat{\mathcal{A}}_n^+$ is the loop algebra associated to $\mathcal{A}_n^+$, algebra $\hat{\mathcal{A}}_n^0$ is the affine Kac-Moody extension of $\mathcal{A}_n^0$ with non-vanishing central extensions. Algebra $\hat{\mathcal{A}}_n^{+}$ is an ideal of the semi-direct sum $\hat{\mathcal{A}}_n^{+} \oplus _s \hat{\mathcal{A}}_n^{0}$; the latter is not semi-simple. The semi-direct sum $\mathcal{A}^{+}_n \oplus _s \mathcal{A}^{0}_n$ is included in the Borel subalgebra of $sl(n)$, coinciding with the latter in the case $n=2$. This means that the $m$-deformed CFTs defined by (\ref{nuevaaccion}) may exhibit an infinite-dimensional symmetry apart from local conformal invariance. 

One could still raise the question as to whether the full $W_n$ symmetry is actually broken. That is, as it happens with conformal symmetry, which is preserved after the introduction of the $\beta $-$\gamma $ system and the shifting $\delta Q$ in the background charge, one may wonder whether the $W^{n>2}(z)$ currents do not suffer from similar modifications and still represent a symmetry of the theory. A naive attempt to construct such modified currents would be shifting the background charge contribution in the $W^3(z)$ current of the $sl(3)$ TFT and adding to it a piece
\begin{equation}
W^3_{\beta \gamma}(w) = \frac{1}{\sqrt{6}} (\partial \beta_1 \partial \gamma_1 - \beta_1 \partial^2\gamma_1 + \partial \beta_2 \partial \gamma_2 - \beta_2 \partial^2\gamma_2),
\end{equation}
which indeed yields
\begin{equation}
W^3_{\beta \gamma}(z)W^3_{\beta \gamma}(w) =\frac{4/3}{(z-w)^6} + \frac{2W_{\beta \gamma}^2(w)}{(z-w)^4} + \frac{\partial W^2_{\beta \gamma}(w)}{(z-w)^3} + ...
\end{equation}
where the ellipses stand for quadratic and simple poles, and where 
\begin{equation}
W_{\beta \gamma}^2(z)= \beta_1\partial\gamma_1 + \beta_2\partial\gamma_2 
\end{equation} 
is the correct contribution of the ghost system to the stress-tensor. However, this direct sum proposal can be seen not to work, the reason being the non-linear nature of the $W_3$ algebra. To the best of our knowledge, there is no evident systematic manner to deform the $W^{n>2}$ currents and find $W$-symmetry in the $m$-deformed CFT. At least in the case of $W_3$ the question about whether such enhanced $W$-symmetry exists is motivated by the fact that the theory seems to have {\it too many} fields for a current algebra such as $\hat{\mathcal A}_3 \oplus \hat{\mathcal A}_3$. One would expect the CFT to be well defined -at least in the sense of the conformal bootstrap- if it has enough symmetry generators\footnote{G.G. thanks Sylvain Ribault for conversations about this point.}, and therefore it is certainly an interesting question whether the model (\ref{nuevaaccion}) exhibits larger symmetry.

A related question is the following: Since the $m$-deformed CFT seems to break the original $W_n$ symmetry to Virasoro symmetry, it would be fully defined only after a complete list of Virasoro primaries together with an algorithm to compute their 3-point point functions are provided. The question arises as to whether the Virasoro primaries considered here form a complete basis or, at least, a sector closed. While equation (\ref{Gmoco}) provides a collection of such primaries whose correlation functions are defined in terms of the TFT observables (\ref{relacion}), the spectrum of Virasoro primary operators could be a priori larger, since in the undeformed theory the $W_n$ module can be decomposed in multiple $W_2$ modules. The situation would be somehow more problematic if the theory happens to exhibit full $W$-symmetry, as in that case it is not sufficient to consider only $W_n>2$ primaries to fully solve the CFT \cite{SexPistols}. This is precisely why providing techniques alternative to the bootstrap, such as the path integral techniques of \cite{Schomerus} is important, specially in the case of non-rational CFTs. 

\subsection{Hidden Kac-Moody symmetry}

As suggested in \cite{Babaro}, the existence of this hidden symmetry generated by $\hat{{\mathcal A}}_n\oplus \hat{{\mathcal A}}_n$, which is a subalgebra of $\hat{sl}(3)\oplus \hat{sl}(3)$, invites to look for a generalization of the deformation (\ref{nuevaaccion}) that, if supplemented with the additional fields in order to realize the additional generators, happens to exhibit full $\hat{sl}(3) \oplus \hat{sl}(3)$ affine Kac-Moody symmetry. In order to look for such a theory, let us consider the action
\begin{equation}\begin{split}
S_{\{ m_k; {\delta}\}}\ =\ &\frac{1}{2\pi }\int d^{2}w
\bigg( (\partial \phi, \bar{\partial}\phi) \,+\,\sum_{k=1}^{2}\left(\beta_k\bar{\partial}\gamma_k\,+\,%
\bar{\beta}_k\partial \bar{\gamma}_k\right)\,+\,\frac{(Q_{T}+\delta Q,\phi)}{4}\mathcal{R}\,+\\&
(-1)^{m_1}\left(\beta_1- \delta \right)^{m_1}\left(\bar{\beta}_1- \bar{\delta } \right)^{m_1}\,e^{\sqrt{2}b(e_1,
\phi)}\,+ (-\beta_2\bar{\beta}_2)^{m_2}\,e^{\sqrt{2}b(e_2,\phi)}\bigg) , \label{HHH}
\end{split}\end{equation}
which is a deformation similar to the one considered before for the case $n=3$ that, apart from the kind of deformation of the type (\ref{nuevaaccion}), also includes a shifting in the ghost field $\beta_1$.

The theory defined by action (\ref{HHH}) may represent a conformal field theory exhibiting a larger algebra that the one generated by the current (\ref{nuevascoientes}) above, provided an adequate relation between $\delta $ and the fields of the theory is prescribed. For instance, if one introduces a third copy of the $\beta $-$\gamma $ system, by adding to (\ref{HHH}) a piece
\begin{equation}
S_{\beta_3 \gamma_3 }=\frac{1}{2\pi }\int d^2z \ \left( \beta_3\bar{\partial }\gamma_3 + \bar{\beta}_3{\partial }\bar{\gamma}_3  \right) , \label{HJK}
\end{equation}
and considers the deformation $\delta  =  -\gamma_2\beta_3 $, making the fields (\ref{HJK}) to interact, and chooses $m_1=m_2=1$, for which $Q_{T}+\delta Q=b\rho$, then one finds that the action (\ref{HHH}) exhibits a hidden full $\hat{sl}(3)_k\oplus \hat{sl}(3)_k$ affine symmetry with Kac-Moody level $k=b^{-2}+3$ \cite{BershadskyOoguri, Bershadsky}. To see this explicitly, one writes down the $\hat{sl}(3)$ currents
\begin{eqnarray}
  J_1^+ (z)&=& \beta_1 \ , \ \ \ \  J_1^0 (z)= {\sqrt{2}}{b}^{-1}(e_1,\partial\phi)-2\gamma_1\beta_1+\gamma_2\beta_2-\gamma_3\beta_3  \nonumber \\
  J_2^+ (z)&=& \beta_2+\gamma_1\beta_3 \ , \ \  J_2^0 (z)= {\sqrt{2}}{b}^{-1}(e_2,\partial\phi)+\gamma_1\beta_1-2\gamma_2\beta_2-\gamma_3\beta_3\nonumber
\end{eqnarray}  
together with $J_3^+ (z)= \beta_3$ and
\begin{eqnarray}
  J_1^- (z)&=& {\sqrt{2}}{b}^{-1}(e_1,\partial\phi)\gamma_1-k\partial\gamma_1-\gamma_3\beta_2-\gamma_1\gamma_1\beta_1
  +\gamma_1\gamma_2\beta_2-\gamma_1\gamma_3\beta_3\nonumber \\
  J_2^- (z)&=& {\sqrt{2}}{b}^{-1}(e_2,\partial\phi)\gamma_2-(k-1)\partial\gamma_2+\gamma_3\beta_1-\gamma_2
  \gamma_2\beta_2\nonumber \\
  J_3^- (z)&=& {\sqrt{2}}{b}^{-1}(e_1,\partial\phi)\gamma_3
	+{\sqrt{2}}{b}^{-1}(e_3,\partial\phi)\gamma_3
	-{\sqrt{2}}{b}^{-1}(e_2,\partial\phi)\gamma_1\gamma_2-k\partial
  \gamma_3+\nonumber \\
	&&(k-1)\gamma_1\partial\gamma_2-\gamma_1\gamma_3\beta_1\,-\,\gamma_2\gamma_3\beta_2
  -\gamma_3\gamma_3\beta_3-\gamma_1\gamma_2\gamma_2\beta_2 ,
\end{eqnarray}
with $b^{-2}=k-3$, and with the free field correlators $\langle \phi_{k} (z)\phi_{\ell } (w)\rangle \sim -({1}/{2})\delta_{k,\ell }\log (z-w)$ and $\langle \beta_i(z)\gamma_j(w)\rangle \sim {\delta_{i,j}}/{(z-w)}$, now with $k,\ell =1,2$ and $i,j = 1,2,3$. It is possible to verify that the OPE between these eight currents and the interaction operators
\begin{equation}
\tilde{S}^1 = (\beta_1 +\gamma_2 \beta _3) (\bar{\beta}_1 +\bar{\gamma}_2 \bar{\beta}_3) e^{\sqrt{2}b(e_1,\phi)} \ ,  \ \ \ \ \ 
{S}^2 = \beta _2 \bar{\beta}_2  e^{\sqrt{2}b(e_2,\phi)}
\end{equation}
is regular, up to total derivatives. Notice that, excluding the contribution of the third ghost system (\ref{HJK}), the currents $J_1^0$, $J_2^0$, $J_1^+$, and $J_2^+$ above coincide with the currents (\ref{nuevascorrientes}) in the case $m_1=m_2=1$.  

Other deformations of this type, such as $m_1=m_2=b^2$, also enjoy full $\hat{sl}(3)_{\hat{k}}\oplus \hat{sl}(3)_{\hat{k}}$ symmetry, in such case with $\hat{k}=(3k-8)/(k-3)$. However, the case $m_1=m_2=1$ is special: in this case, action (\ref{HHH}) augmented with the system (\ref{HJK}) actually corresponds to the $SL(3,\mathbb{R})$ WZW model at level $k=b^{-2}+3$ written in Wakimoto variables. In other words, $S_{\text{ WZW}} = S_{\{ m_{1,2}=1;\delta =-{\gamma_2 \beta_3 } \}} + S_{\beta_3 \gamma_3 }$. The central charge in this case is given by $c_{\text{WZW}}= 8+24b^{2}=8k/(k-3)$. The relation between the level $k$ and the parameter $b$ of the undeformed TFT is the same as in the Drinfeld-Sokolov Hamiltonian reduction \cite{Drinfeld-Sokolov, BershadskyOoguri, Bershadsky}. To see this explicitly, consider the WZW action
\be
S_{\text{WZW}}[g]=\f{k}{2\pi}\int_{S} d^2 z \, \text{Tr} \big{(}g^{-1}\partial g\, g^{-1}\bar{\partial} g \big{)}+ 
\f{k}{12\pi}\int_{B} d^3{x} \, \epsilon^{\nu\sigma\kappa}\, \text{Tr}\big{(} \hat{g}^{-1}\partial_\nu \hat{g}\,  \hat{g}^{-1}\partial_\sigma \hat{g}\, \hat{g}^{-1}\partial_\kappa \hat{g} \big{)} \label{WZW56}
\ee
where $k$ is the WZW level, $g(z)$ is a group valued field on $S$, $g\in SL(3,\mathbb{R})$; $S$ is a 2-dimensional surface that coincides with the boundary of $B$, i.e. $S=\partial B$. $\hat{g}(x)$ is the extension of $g(z)$ in the 3-dimensional ambient $B$. To parameterize the group element, consider the Jordan-Chevalley decomposition
\begin{equation}
g=e^{-\gamma_1 T_1^- -\gamma_2 T_2^- -(\gamma_3-\frac{1}{2}\gamma_1 \gamma_2) T^-_3} e^{\phi_1 \tilde{T}^0_1+\phi_2 \tilde{T}^0_2} e^{-\bar{\gamma}_1 T^+_1-\bar{\gamma}_2 T^+_2-(\bar{\gamma}_3-\frac{1}{2}\bar{\gamma}_1\bar{\gamma}_2) T^+_3},
\end{equation}
with $T_i^A$ being the generators of $sl(3)$ given by the upper triangular matrices
\bea
T^+_1=\begin{pmatrix} 0 & 1 & 0 \\ 0 & 0 & 0 \\ 0 & 0 & 0 \end{pmatrix},\qquad 
T^+_2&=&\begin{pmatrix} 0 & 0 & 0 \\ 0 & 0 & 1 \\ 0 & 0 & 0 \end{pmatrix},\qquad 
T^+_3=\begin{pmatrix} 0 & 0 & 1 \\ 0 & 0 & 0 \\ 0 & 0 & 0 \end{pmatrix},\nonumber
\eea
together with the lower triangular
\bea
T^-_1=\begin{pmatrix} 0 & 0 & 0 \\ 1 & 0 & 0 \\ 0 & 0 & 0 \end{pmatrix},\qquad 
T^-_2&=&\begin{pmatrix} 0 & 0 & 0 \\ 0 & 0 & 0 \\ 0 & 1 & 0 \end{pmatrix},\qquad 
T^-_3=\begin{pmatrix} 0 & 0 & 0 \\ 0 & 0 & 0 \\ 1 & 0 & 0 \end{pmatrix},\nonumber
\eea
and the two Cartan elements
\bea
T^0_1&=&\begin{pmatrix} 1 & 0 & 0 \\ 0 & -1 & 0 \\ 0 & 0 & 0 \end{pmatrix},\qquad 
T^0_2=\begin{pmatrix} 0 & 0 & 0 \\ 0 & 1 & 0 \\ 0 & 0 & -1 \end{pmatrix};\nonumber
\eea
it is convenient to define the basis $\tilde{T}^0_1=({T^0_1+T^0_2})/{\sqrt{2}}$, $\tilde{T}^0_2=({T^0_1-T^0_2})/{\sqrt{6}}$. The holomorphic conserved currents that generate the $\hat{sl}(3)$ affine algebra are
\begin{equation}
J_{i}^A(z)=\text{Tr}(J(z) T^A_i )  \ , \ \ \ \ \ \ \ J(z) = k\partial g g^{-1} \label{currento}
\end{equation}
with $i=1,2,3$ for $A=\pm $ and $i=1,2$ for $A=0$. The anti-holomorphic currents can be written in a similar manner, with $\bar{J}(z) = -kg^{-1}\bar{\partial }g $. The simple roots of $sl(3)$ are
\bea
{e}_1=\frac{1}{\sqrt{2}}\left(1,\sqrt{3}\right),\qquad
{e}_2=\frac{1}{\sqrt{2}}\left(1,-\sqrt{3}\right),
\eea
with fundamental weights 
\be
{\omega}_1=\frac{1}{\sqrt{2}}\left(1,\frac{1}{\sqrt{3}}\right),\qquad
{\omega}_2=\frac{1}{\sqrt{2}}\left(1,-\frac{1}{\sqrt{3}}\right) , 
\ee
from which one easily verifies $(\omega_i ,e_j)=\delta_{ij}$ and $(e_1,e_1)=(e_2,e_2)=2$, $(e_1,e_2)=(e_2,e_1)=-1$; the Weyl vector reads $\rho = \omega_1+\omega_2=(\sqrt{2},0)$. Plugging this representation in (\ref{currento}), defining the fields $\beta_i$
\bea
\beta_1&=&-k e^{\sqrt{2}b(e_1,\phi)}\partial \bar{\gamma}_1 -\gamma_2 \beta_3,\nonumber\\
\beta_2&=&-k e^{\sqrt{2}b(e_2,\phi)}\partial \bar{\gamma}_2,\nonumber\\
\beta_3&=&-k e^{\sqrt{2}b(\rho,\phi)} \left(\partial\bar{\gamma}_3-\bar{\gamma}_2 \partial \bar{\gamma}_1\right),
\eea
(similarly for $\bar{\beta }_i$), and taking into account quantum corrections that amount to renormalize $\phi \to \sqrt{2} b \phi $, one eventually verifies that the $SL(3,\mathbb{R})$ WZW action (\ref{WZW56}) takes the form\footnote{Where we also shifted the zero-modes of the fields $\phi_1$ and $\phi_2$ in order to absorb an overall factor $b^{-2}$ in the interaction terms.}
\bea
S_{\text{WZW}}&=&\f{1}{2\pi}\int d^2 z \Bigg( (\partial \phi, \bar{\partial} \phi)+\sum_{i=1}^{3}(\beta_i \bar{\partial} \gamma_i+\bar{\beta}_i \partial \bar{\gamma}_i)+\f{b(\rho ,\phi )}{4} {\mathcal R} - \nonumber\\
&& (\beta_1+\beta_3\gamma_2)(\bar{\beta}_1+\bar{\beta}_3\bar{\gamma}_2) e^{\sqrt{2}b(e_1,\phi)}-\beta_2 \bar{\beta}_2 e^{\sqrt{2}b(e_2,\phi)}+ \text{c.t.} \Bigg). \label{cosito}
\eea
where $\text{c.t.}$ stands for a contact term; more precisely, for a term 
\begin{equation}
\text{c.t.} = -\frac{1}{2\pi}\int d^2z \beta_3 \bar{\beta}_3 e^{\sqrt{2}b(\rho,\phi)} . \label{contacto} 
\end{equation}
In Ref. \cite{Hikida-sl3}, a different parameterization of the $SL(3,\mathbb{R})$ elements is considered. Such parameterization leads to a more symmetric form of the action (\ref{cosito}), introducing a shift $\delta $ also in the last term what makes the two screening operators to look similar. The parameterization we considered here has some advantage for the Coulomb gas computation. The contact term in \cite{Hikida-sl3} takes, however, exactly the same form as the one here, namely (\ref{contacto}). The same interpretation for such term as the one given in \cite{Hikida-sl3} holds here. Then, we observe that, up contact terms, action (\ref{cosito}) actually agrees with $S_{\{ m_{1,2}=1;\delta =-{\gamma_2 \beta_3 } \}} + S_{\beta_3 \gamma_3 }$.

\section{Discussion}

In this paper, we have constructed an infinite-dimensional family of two-dimensional conformal field theories that admit Lagrangian representation. These theories consist of particular deformations of $sl(n)$ Toda field theories. Such deformations preserve conformal invariance and deform the full $W_n$ symmetry. As a recognition for having being indulgent with the conformal symmetry, we have been left with a remnant infinite-dimensional symmetry $\hat{\mathcal A}_n\oplus \hat{\mathcal A}_n$ which, in the particular case $n=3$, can be enhanced to full $\hat{sl}(3) \oplus \hat{sl}(3)$ affine symmetry if other deformation operators are allowed. We have explored here the simplest cases of deformation of TFT, which basically consist in the most direct extension of the results of \cite{Ribault-family} to the $sl(n)$ case with $n\geq 2$. It would be interesting to explore other types of deformations and their possible physical applications. Some open questions regarding this are the following: First, whether a systematic way of deforming $sl(n)$ TFT is possible such that one obtains a full $\hat{sl}(n)\oplus \hat{sl}(n)$ affine symmetry for $n>3$. This would extend the WZW-Liouville correspondence to higher rank and for more general $sl(2)$ embeddings. Secondly, it would be interesting to have a full understanding of the relation between the Hamiltonian reduction at quantum level and the correspondence between correlators of theories with $W_n$ symmetry and of theories with $\hat{sl}(n)$ affine symmetry. Interesting results in this direction have been obtained recently in \cite{Hikida-sl3}. A third question that remains open is the aforementioned problem of proving whether or not the deformed theory exhibits hidden $W_n$ symmetry. It appears to us that there is no obvious, systematic way of showing this. This could be seen as an obstruction, since the theory seems not to have a symmetry algebra as large as needed to be solved by bootstrap methods. This is precisely why alternative techniques such as the path integral approach of \cite{Schomerus} are important, in particular when dealing with non-rational CFTs. 

Lastly, it would be interesting to see whether there exists a concrete application of the deformed TFT to study gauge theories via AGT and its generalization. It turns out that the $m$-deformed theories do offer a potential application within this context, which is the description of defects in the ${\mathcal N}=2$ superconformal $SU(n)$ quiver theories: According to the Wyllard's generalization of AGT conjecture, the Nekrasov partition function of such $SU(n)$ theories is in correspondence with $sl(n)$ TFT correlation functions. In the case $n=2$, it is known how to generalize the correspondence in order to describe not only the partition function but also expectation values of a whole set of surface and loop operators in the gauge theory side \cite{gordos1, gordos2}. Such observables are also given by Liouville correlation functions, but including degenerate fields, namely fields that contain null Virasoro descendants. The vacua of surface operators in the gauge theory are labeled by integer numbers that are in correspondence with the level of the null vectors in the 2-dimensional CFT. Non-fundamental surface operators of this type are believed to exist in generic $SU(n)$ ${\mathcal N}=2$ gauge theories too, and the expectation value of such operators would also admit a 2-dimensional CFT description in terms of TFT observables. TFT contains degenerate and semi-degenerate representations, and the possibility of the correlators of the theory (\ref{nuevaaccion}) for the appropriate values of $\{ m_1 , m_2 , ... ,m_{n-1} \}$ describing expectation values of non-fundamental operators in $SU(n)$ gauge theories in certainly interesting. This idea has been discussed in \cite{Giribet-Babaro2} for the case $n=2$, where it was argued that defects in the ${\mathcal N}=2^*$ $SU(2)$ gauge theory could be associated to a theory with affine symmetry. The affine CFT description of surface operators in ${\mathcal N}=2$ theories was suggested in \cite{AT}. Another possibly related result is that of reference \cite{ultimatum}, where, based on previous results \cite{ultimatum2} for the TFT correlation functions, it was argued how the inclusion of a semi-degenerate primary operator in the TFT 3-point function corresponds in the gauge theory side to a particular Higgsing of the non-Lagrangian theory on $S^4$. It would be interesting to make these ideas precise and study the potential applications of these deformed theories within the context of the 2D/4D correspondence.

\[
\]

The authors thank Simone Giacomelli and Sylvain Ribault for interesting remarks. This work has been partially funded by FNRS-Belgium (convention FRFC PDR T.1025.14 and convention IISN 4.4503.15), by CONICET of Argentina, by the Communaut\'{e} Fran\c{c}aise de Belgique through the ARC program and by a donation from the Solvay family. The Centro de Estudios Cient\'{\i}ficos (CECs) is funded by the Chilean Government through the Centers of Excellence Base Financing Program of CONICYT-Chile.


\end{document}